\documentclass[a4paper,11pt]{article}
\usepackage{amsmath}
\usepackage{amsfonts}
\usepackage{amssymb}
\usepackage[utf8x]{inputenc}
\usepackage{ae}
\usepackage[T1]{fontenc}
\usepackage[english]{babel}
\usepackage{graphicx}
\usepackage{placeins}
\usepackage{tabularx}
\usepackage{tabulary}
\usepackage{setspace}
\usepackage{tablefootnote}
\usepackage[flushmargin]{footmisc}
\usepackage[comma]{natbib}
\usepackage{array}
\usepackage{float}
\usepackage{caption}
\usepackage[a4paper,left=2.5cm,right=2.5cm,top=3cm,bottom=3cm]{geometry}
\usepackage[colorinlistoftodos]{todonotes}
\usepackage{pdfpages}
\usepackage{longtable}
\usepackage{adjustbox}
\usepackage{booktabs}
\usepackage{lscape}
\usepackage[colorlinks=true, allcolors=blue]{hyperref}

\usepackage[flushleft]{threeparttable}
\usepackage{inputenc}
\usepackage{soul}
\usepackage{lineno}



\begin{document} \doublespacing \pagestyle{plain}
	
	\def\ci{\perp\!\!\!\perp}
	\begin{center}
		
		{\Large Business analytics meets artificial intelligence:\\Assessing the demand effects of discounts on Swiss train tickets}
		
		{ \vspace{0.4cm}}
		
		{ Martin Huber*, Jonas Meier**, and Hannes Wallimann+ }\medskip
		
		{\small {*University of Fribourg, Dept.\ of Economics
\\ **University of Amsterdam, Faculty of Economics and Business\\+University of Applied Sciences and Arts Lucerne, Competence Center for Mobility} \bigskip }
	\end{center}
	
	\smallskip

	\noindent \textbf{Abstract:} {We assess the demand effects of discounts on train tickets issued by the Swiss Federal Railways, the so-called `supersaver tickets', based on machine learning, a subfield of artificial intelligence. Considering a survey-based sample of buyers of supersaver tickets, we use causal machine learning to assess the impact of the discount rate on rescheduling a trip, which seems relevant in the light of capacity constraints at rush hours. Assuming that (i) the discount rate is quasi-random conditional on our rich set of characteristics and (ii) the buying decision increases weakly monotonically in the discount rate, we identify the discount rate's effect among `always buyers', who would have traveled even without a discount, based on our survey that asks about customer behavior in the absence of discounts. We find that on average, increasing the discount rate by one percentage point increases the share of rescheduled trips by 0.16 percentage points among always buyers. Investigating effect heterogeneity across observables suggests that the effects are higher for leisure travelers and during peak hours when controlling several other characteristics.}
	
	
	{\small \smallskip }
	
	

	{\small \noindent \textbf{Acknowledgments:}  We are grateful to the SBB Research Fund for financial support. Furthermore, we are indebted to Pierre Chevalier, Reto Lüscher, Philipp Wegelin and Kevin Blättler for their helpful discussions. \textbf{Addresses for correspondence:} Martin Huber, University of Fribourg, Bd.\ de P\'{e}rolles 90, 1700 Fribourg, Switzerland, martin.huber@unifr.ch; Jonas Meier, University of Amsterdam, Roeterstraat 11, 1018WB Amsterdam, Netherlands, j.c.meier@uva.nl; Hannes Wallimann, University of Applied Sciences and Arts Lucerne, Rösslimatte 48, 6002 Luzern, Switzerland, hannes.wallimann@hslu.ch.
	}\thispagestyle{empty}\pagebreak

	{\small \renewcommand{\thefootnote}{\arabic{footnote}} %
		\setcounter{footnote}{0}  \pagebreak \setcounter{footnote}{0} \pagebreak %
		\setcounter{page}{1} }
	
	\section{Introduction}\label{intro}
	
	Organizing public transport involves a well-known trade-off between consumer welfare and provider revenue. Typically, consumers value frequency, reliability, space, and low fares \citep{Redman+2013} while suppliers aim at operating with a minimum number of vehicles to maximize profits. In general, the allocation can be improved as providers do not account for the positive externalities on consumers \citep{mohring1972optimization}. In particular, service frequency reduces travelers' access and waiting costs. This so-called `Mohring-effect' leads to economies of scale, implying the need for subsidies to achieve the first-best solution in terms of welfare. Consequently, it may be socially optimal to subsidize railway companies to reduce fares \citep[]{ParrySmall2009}. To assess such a measure's effectiveness on demand, policy-makers would need to know how individuals respond to lower fares.  However, it is generally challenging to identify causal effects of discounts on train tickets (or goods and services in general) due confounding or selection. For instance, discounts might typically be provided for dates or hours with low train utilization such that connections with and without discount are not comparable in terms of baseline demand. A naive comparison of sold tickets with and without discount would therefore mix the influence of the discount with that of baseline demand. In this context, we apply machine learning (a subfield of artificial intelligence) to convincingly assess how discounts on train tickets for long-distance connections in Switzerland, the so-called `supersaver tickets', affect demand, by exploiting a unique data set of the Swiss Federal Railways (SBB) that combines train utilization records with a survey of supersaver buyers.
	
	More concisely, our study provides a promising use case of causal machine learning for business analytics in the railway industry: analysing the causal effect of the discount on demand shifts among customers that would have booked the trip even without discount. This is feasible because our unique survey contains information on how supersaver buyers would have decided in the absence of a discount, e.g. whether they are so-called `always buyers' and would have booked the connection even at the regular fare. Using causal machine learning approaches, this novel survey allows us to answer the question whether discounts causally affect individual buying decisions - an important and under-researched question. Moreover, and to the best of our knowledge, this is the first application of causal machine learning techniques in the public transportation literature.
	
	For both our descriptive and causal analysis, we make use of appropriately tailored machine learning techniques, which learns the associations between the demand outcomes of interest, the discount rate, and further customer or trip-related characteristics in a data-driven way and helps avoiding model misspecification. Predictive machine learning permits optimizing forecasts about demand and customer behavior as a function of observed characteristics in order to descriptively  analyse specific patterns in customer behavior. On the other hand, causal machine learning permits evaluating the causal effect of specific interventions like a discount regime for optimizing the offer of such discounts. When applying predictive machine learning, trip-related characteristics like seat capacity, utilization, departure time, and the discount rate, but also customer's age turn out to be strong predictors.
	
	
	
	Concerning the causal analysis, we impose (i) a selection-on-observables assumption implying that the discount rate is as good as randomly assigned when controlling for our rich set of trip- and demand-related characteristics and (ii) weak monotonicity of any individual's decision to purchase an additional trip (otherwise not realized) in the discount rate, implying that a higher (rather than lower) discount does either positively or not affect any customer's buying decision. We formally show how these assumptions permit tackling the selectivity of discount rates and survey response to identify the discount rate's effect on demand shifts (rescheduling away from rush hours) for the subgroup `always buyers', based on the survey information on how customers would have behaved in the absence of a discount. Furthermore, we discuss testable implications of monotonicity, namely that among all survey respondents, the share of additional trips must increase in the discount rate, and the selection on observables assumptions, requiring that conditional on trip- and demand-related characteristics, the discount must not be associated with personal characteristics (like age or gender) among always buyers. Hypothesis tests do not point to the violation of these implications. 	
	
	Based on our causal identification strategy, we estimate the marginal effect of slightly increasing the (continuously distributed) discount rate based on the causal forest (CF), see \cite{WagerAthey2018} and \cite{AtheyTibshiraniWager2019}, and find that on average, increasing the discount rate by one percentage point increases the share of rescheduled trips by 0.16 percentage points among always buyers. In a second approach, we binarize the discount rates by splitting them into two discount categories of less than 30\% (relative to the regular fare) and 30\% or more. Applying double machine learning (DML), see \cite{Chetal2018}, we find that discount rates of 30\% and more on average increase the share of rescheduled trips 3.6 percentage points, which is in line with the CF-based results. Our paper therefore provides the first empirical evidence (at least for Switzerland)  that such discounts can help balancing out train utilization across time and reducing overload during peak hours, albeit the magnitude of the impact on always buyers appears limited.\footnote{We note that the CF and DML estimators are but two approaches in the emerging and fast developing literature on combining causal inference and machine learning, see for instance causal bayesian networks  \citep[e.g. ][]{heckerman2013bayesian} or probabilistic graphical models \citep[e.g. ][]{sucar2015probabilistic} for further examples. For surveys on recent developments in causal analysis based on machine learning and algorithms, we refer the reader to  \cite{AtheyImbens2019} and \cite{peters2017elements}.} We benchmark our estimated effects based on linear regression and propensity score matching. The findings based on the latter approaches point in the same direction as our main results. Yet, the results become less significant and for propensity score matching, this goes together with an increase in the variance. This may be due to overfitting, given our relatively high number of potential control variables. Therefore, in similar scenarios as ours, causal machine learning may appear preferable, as the data-driven selection of important control variables (rather than using any of them) reduces the risk of overfitting. Finally, our study also highlights the advantages of flexibly including influential control variables  among many observed covariates - a task for which causal machine learning is well suited.
	
	When investigating the heterogeneity of effects across all of our observed characteristics using the CF, our results suggest that demand-related trip characteristics (like seat capacity, utilization, departure time, and distance) have some predictive power for the size of the discounts' impact on shifting demand. Such information on heterogeneous effects appears interesting for optimizing the allocation of discounts for the purpose of shifting demand, as the SBB has (due to its monopoly in the Swiss long-distance passenger rail market) agreed with the Swiss price monitoring agency to provide a fixed amount of discounted tickets per year, but is free to chose the timing and connections for discounts. In a second heterogeneity analysis, we investigate whether effects differ systematically across a pre-selected set of characteristics, namely: age, gender, possession of a half fare travel card, travel distance, whether the purpose is business, commute, or leisure, and whether the departure time is during peak hours. Using the regression approach of \cite{SemenovaChernozhukov2020}, we find that conditional on the other characteristics, the effects of increasing the discount by one percentage point on rescheduling are by more than 0.2 percentage points higher during peak hours and for leisure travelers, differences that are statistically significant at the 10\% level when, however, not controlling for multile hypothesis testing. These effects appear plausible as leisure travelers are likely more flexible and discounts during peak hours make trips at times of increased demand even more attractive. We do not find statistically significant effect differences for the other pre-selected characteristics, which could, however, be due to the (for the purpose of investigating effect heterogeneity) limited sample of several thousand observations.
	
	Our paper is related to a growing literature applying statistical and machine learning methods for analyzing transport systems, as well as to methodological studies on causal inference for so-called principal strata, see \cite{FrangakisRubin02}, i.e. endogenous subgroups like the always buyers. Typically, it is hard to identify the causal effect of some treatment (or intervention) like a discount on such a non-randomly selected subgroup defined in terms how a post-treatment variable (e.g.\ buying decision) depends on the treatment (e.g.\ treatment). One approach is to give up on point identification and instead derive upper and lower bounds on a set of possible effects for groups alike the always buyers based on the aforementioned monotonicity assumption (and possibly further assumptions about the ordering of outcomes of always buyers and other individuals), see for instance \cite{ZhRu03}, \cite{ZhRuMe08}, \citet{Im08}, \citet{Le05}, and \citet{blanco2011bounds}. Alternatively, the treatment effect on always buyers is point-identified  when invoking a selection-on-observables or instrumental variable assumption for selection into the survey, see for instance \cite{Hu11b}, which requires sufficiently rich data on both survey participants and non-participants for modeling survey participation. In contrast to these previous studies, the approach in this paper point-identifies the treatment effect by exploiting the rather unique survey feature that customers were asked about their behavior in the absence of the discount, which under monotonicity permits identifying the principal stratum of always buyers directly in the data.
	
	Furthermore, our work is related to conceptual studies on transport systems, considering for instance the previously mentioned positive externalities of an increased service for customers that are not accounted for by transportation providers. Such externalities typically arise from economies of scale due to fixed costs and a 'Mohring effect', implying that service frequency reduces waiting costs \citep{mohring1972optimization}. The study by \cite{ParrySmall2009} suggests that lower fares can boost overall welfare by increasing economies of scale (off-peak) and decreasing pollution and accidents (at peaks). Similarly, \cite{Palma+2017} argue that time-dependent ticket prices may increase overall welfare as overcrowding during peak hours is suboptimal for both consumers and providers. As public transport is usually highly subsidized, governments may directly manage the trade-off mentioned above. As this involves taxpayer money, it is a question of general interest how the subsidies should be designed. Based on their results, \cite{ParrySmall2009} conclude  that even substantial subsidies are justified due to lower fares' positive welfare effect. In contrast, \cite{BassoSilva2014} find that the contribution of transit subsidies to welfare diminishes once congestion is taxed and alternatives are available, i.e., bus lanes. Irrespective of the specific policy instrument, the consumer's willingness to shift demand drives these policies' effectiveness. While many factors affect this willingness, most studies conclude that consumers are price sensitive \citep{Paulley+2006}. In this context, we aim at contributing to a better understanding of how time-dependent pricing translates to consumer decisions. In particular, we shed light on the effects of a continuously distributed discount. The recent literature typically considers binary or categorical treatments \citep[e.g.,][]{kholodov2021public,shin2021exploring,wallimann2021price}, whereas studies on continuous treatments are rare in the public transportation literature.
	
	More broadly, our paper relates to the literature on policies targeting demand shifts. Among these, the setting of car parking costs, fiscal regulations, or even free public transport has been analyzed \citep[e.g.][]{batty2015challenges, RotarisDanielis2014, Zhang+2018,de2006impact}. Another stream of literature applies machine learning algorithms in the context of public transport. Examples are short-term traffic flow forecasts for bus rapid transit \citep[][]{liu2017novel} or metro \citep[][]{liu2019deeppf} services. Further, \cite{hagenauer2017comparative} and \cite{omrani2015predicting} implement machine learning algorithms to predict travel mode choices. \cite{yap2020predicting} predict disruptions and their passenger delay impacts for public transport stops. In other research fields, also applications of causal (rather than predictive) machine learning are on the rise \citep[see for instance][]{yang2020double,knaus1805double}. This is, to the best of our knowledge, the first study using causal machine learning in the context of public transport. Finally, a growing literature discusses the opportunities of data-driven business decision-making \citep[][]{brynjolfsson2016rapid} by assessing the relevance of predictive and causal machine learning. \cite{ascarza2018retention} and \cite{hunermund2021causal} show that companies may gain by designing their policies based on causal machine learning. For instance, firms can target the relevant consumers much more effectively when accounting for their heterogeneity in terms of reaction to a treatment. Our study provides a use case of how the machine learning-based assessment of discounts could be implemented also in other businesses and industries facing capacity constraints. 
	
	This paper proceeds as follows. Section \ref{Background} presents the institutional setting of passenger  railway transport in Switzerland. Section \ref{data} describes our data, coming from a unique combination of a customer survey and transport utilization data. Section \ref{ident} discusses the identifying assumptions underlying the causal machine learning approach as well as testable implications. Section \ref{methods} outlines the employed causal machine learning methods. Section \ref{descriptives} contains a descriptive analysis of the data, including predictive machine learning. Section \ref{resuls} presents the causal results. Finally, Section \ref{conclusion} concludes.
	
	\section{Institutional background}\label{Background}
	The railway system in Switzerland is known for its high quality of service. Examples include the high level of system integration with frequent services, synchronized timetables, and comprehensive fare integration, see \cite{desmaris2014reform}. In Switzerland, a country of railway tradition, the state owned incumbent Swiss Federal Railways (SBB) operates the long distance passenger rail market as monopolist \citep[][]{thao2020swiss}. Furthermore, nationally operating long-distance coaches may only be approved if they do not `substantially' compete with existing services. Thus, the SBB competes exclusively with motorized private transport in Swiss long-distance traffic. The company also owns most of the rail infrastructure, which is funded by the Federal Government. However, since the end of 2020 the companies Berne-Lötschberg-Simplon Railways (BLS) and Southeast Railways (SOB) operate a few links on behalf of the SBB. Different to regional public transport that Swiss taxpayers subsidize with approximately CHF 1.9 bn per year, the operation of the long distance public transport itself has to be self-sustaining \citep[][]{wegelin2018mere}.
	
	Because of the monopoly position of the SBB in long distance passenger transport, the prices are screened by the Swiss `price watchdog' (or price monitoring agency) to prevent abuse. Based on the price monitoring act, the watchdog keeps a permanent eye on how prices and profits develop. By the end of 2014, the watchdog concluded that the SBB charged too high prices. As a consequence and through a mutual agreement, the SBB and the Swiss price watchdog agreed on a significantly higher supply of supersaver tickets, which were first offered in 2009. Using a supersaver ticket, customers can travel on long distance public transport routes with a discount of up to 70\%. Thereafter, additional agreements were regularly reached regarding number and scope of the supersaver tickets. While only a few thousand supersaver tickets were sold in 2014, sales increased to about 8.8 million in 2019, see \cite{luescher2020}.
	
	From the SBB's perspective, these tickets can serve two purposes. First, the tickets might be used as means to balance out the utilization of transport services. For instance, supersaver tickets could reduce the high demand during peak hours which is a key challenge for public transport. Thus, balancing the demand may reduce delays and increase the number of free seats which is valued by the consumers. The average load of SBBs’ seats amounts to 30\% in the long distance passenger transport.\footnote{See https://reporting.sbb.ch/verkehr.} For this reason, there is in the literal sense room for improving the allocation. Second, price sensitive customers can be acquired during off-peak hours at rather negligible marginal costs.
	
	Despite the increasing interest in the supersaver tickets in recent years, many users of the Switzerland public transport network purchase a so-called `general abonnement' travel ticket (GA). This (annually renewed) subscription provides free and unlimited access to the public transport network in Switzerland. In 2019, about 0.5 million individuals owned a GA in Switzerland, roughly 6\% of the Swiss population. The GA's cost amounts to 3,860 and 6,300 Swiss francs for the second and first class, respectively. In the same year, about 2.7 million individuals held a relatively cheap half fare travel ticket amounting to 185 Swiss francs. The latter implies a price reduction of 50\% for public transport tickets in Switzerland. Overall, discounts provided through supersaver tickets are slightly lower for owners of half fare tickets, as the SBB aims to attract non-regular public transport users. In our causal analysis, we therefore also control for the possession of a half fare ticket.
	
	\section{Data}\label{data}
	To investigate supersaver tickets' effect, we use a unique cross-sectional data set provided by the SBB. Our sample consists of randomly surveyed buyers of supersaver tickets that purchased their tickets between January 2018 and December 2019. These survey data are matched with data on distances between any two railway stops as well as utilization-related information relevant for the supply and calculation of discounts. In Section \ref{resuls}, we provide descriptive statistics for these data.
	
	\subsection{Survey data}
	The customer survey is our primary data source. It for instance includes the outcome variable `demand shift', a binary indicator of whether an interviewee rescheduled her or his trip due to buying a supersaver ticket. `Yes' means that the departure time has been advanced or postponed because of the discount. A second variable characterizing customer behavior is an indicator for upselling, i.e. whether someone purchased a first rather than a second class ticket as a reaction to the discount. Another question asks whether an interviewee would have bought the train trip in the same or a higher class even without being offered a discount, which permits judging whether an additional trip has been sold through offering the discount and allows identifying the subgroup of always buyers under the assumptions outlined further below. Our continuously distributed treatment variable is the discount rate of a supersaver ticket relative to the standard fare, which may take positive values of up to 70\%.
	
	Furthermore, we observe two kinds of covariates, namely trip- or demand-related factors and personal characteristics of the interviewee. The former are important control variables for our causal identification strategy outlined below and include the difference between the days of purchase and travel, the weekday, month, and year, an indicator for buying a half fare ticket, departure time, peak hour,\footnote{Peak hour is defined as a departure time between 6am and 8:59am or between 4pm and 18:59pm, from Monday to Friday. This time windows is chosen on the base of the SBB's train-path prices. For further details, see \hyperlink{https://company.sbb.ch/en/sbb-as-business-partner/services-rus/onestopshop/services-and-prices/the-train-path-price.html}{https://company.sbb.ch/en/sbb-as-business-partner/services-rus/onestopshop/services-and-prices/the-train-path-price.html} (assessed on March 24 2021).} number of tickets purchased per person, class (first or second), indicators for leisure trips, commutes, or business trips, the number of companions (by children and adults if any) and a judgment of how complicated the ticket purchase was on a scale from 1 (complicated) to 10 (easy). Furthermore, it consists of indicators for the point of departure, destination, and public holidays. The personal characteristics include age,  gender, migrant status, language (German, French, Italian), and indicators for owning a half fare travel ticket or other subscriptions like those of regional tariff associations, specific connections, and Gleis 7 (`rail 7'). The latter is a travelcard for young adults not older than 25, providing free access to public transport after 7pm.

	\subsection{Factors driving the supply of supersaver tickets}
	In addition to the survey, we have access to factors determining the supply of supersaver tickets with various discounts. This is crucial for our causal analysis that hinges on on controlling for all characteristics jointly affecting the the discount rate and the demand shift outcome. While information on the distances between railway stops in Switzerland is publicly available,\footnote{See the Open Data Platform of the SBB: \hyperlink{https://data.sbb.ch/explore/dataset/linie-mitbetriebspunkte}{https://data.sbb.ch/explore/dataset/linie-mitbetriebspunkte} (accessed on March 24 2021), which provides the distances between any stops on a railway route.} the SBB provides us for the various connections with information on utilization data, the number of offered seats, and contingency schemes, which define the quantity of offered discounts. This allows us to account for travel distance, offered seats, capacity utilization, and quantities of offered supersaver tickets for various discount levels as well as quantities of supersaver tickets already sold  (both quantities at the time of purchase). Furthermore, we create binary indicators for the 27 different contingency schemes of the SBB present in our data, which change approximately every month.
	
	The variables listed in the previous paragraph are important, as the SBB calculates the supply of supersaver tickets based on an algorithm considering four type of inputs: Demand forecasts, advance booking deadlines, number of supersaver tickets already sold, and contingency schemes defining the amount and the size of offered discounts based on the three previous inputs. The schemes are set as a function of the SBB’s self-imposed goals such as customer satisfaction but also depend on the requirements imposed by the price watchdog. The algorithm calculates a journey’s final discount as a weighted average of all discounts between any two adjacent railway stops along a journey. The weights depend on the distances of the respective subsections of the trip. To approximate the (not directly available) demand forecasts of the SBB, we consider the quarterly average of capacity utilization and the number of offered seats for any two stops, which are available by (exact) departure time, workday, class, and weekend. In addition, we make use of indicators for place of departure, destination, month, year, weekday and public holidays. We use this information to reconstruct the amount and size of offered discounts by taking values from the contingency schemes that correspond to our demand forecast approximation combined with the difference between buying and travel days. Comparing this amount and size of offered discounts with a buyer’s discount, we estimate the number of supersaver tickets already sold for the exact date of purchase.
	
	\subsection{Sample construction}
	Our initial sample contains 12,966 long-distance train trips that cover 61,469 sections between two adjacent stops. For 12.2\% of these sections, there is no information on the capacity utilization available, which can be due to various reasons. First, for some cases, capacity utilization data is missing. Second, passengers traveling long-distance may switch to regional transport in exceptional cases causing problems for determining utilization. A further reason could be issues in data processing. Altogether, missing information occurs in 3,967 trips of our initial sample. We tackle this problem by dropping all journeys with more than 50\% of missing information, which is the case for 320 trips or 2.5\% of our initial sample. After this step, our evaluation sample consists of 12,646 trips. For the remaining 3,647 trips with missing information (which now account for a maximum of 50\% of all sections of a journey), we impute capacity utilization as the average of the remaining sections of a trip. In our empirical analysis, we include an indicator for whether some trip information has been imputed as well as the share of imputed values for a specific trip as control variables. Finally, we note that our causal analysis makes (in contrast to the predictive analysis) only use of a subsample, namely observations identified as always buyers who would have traveled even without a discount, all in all 6,112 observations.
	
	\section{Identification}\label{ident}
	
	We subsequently formally discuss the identification strategy and assumptions underlying our causal analysis of the discounts among always buyers.
	
	\subsection{Definition of Causal Effects}
	
	Let $D$ denote the continuously distributed treatment `discount rate' and $Y$ the outcome `demand shift', a binary indicator for rescheduling a trip due to being offered a discount. More generally, capital letters represent random variables in our framework, while lower case letters represent specific values of these variables. To define the treatment effects of interest, we make use of the potential outcome framework, see for instance  \cite{Rubin74}. To this end, $Y(d)$ denotes the potential outcome hypothetically realized when the treatment is set to a specific value $d$ in the interval $[0, Q]$, with $0$ indicating no discount and $Q$ indicating the maximum possible discount. For instance, $Q=0.7$ would imply the maximum discount of 70\% of a regular ticket fare. The realized outcome corresponds to the potential outcome under the treatment actually received, i.e.\ $Y=Y(D)$, while the potential outcomes under discounts different to one received remain unknown without further statistical assumptions.
	
	A further complication for causal inference is that our survey data only consists of individuals that purchased a supersaver ticket, a decision that is itself an outcome of the treatment, i.e.\ the size of the discount. Denoting by $S$ a binary indicator for purchasing a supersaver ticket and by $S(d)$ the potential buying decision under discount rate $d$,  this implies that we only observe outcomes $Y$ for individuals with $S=1$. In general, making the survey conditional on buying introduces Heckman-type sample selection (or collider) bias, see \cite{Heck76} and  \cite{He79}, if unobserved characteristics affecting the buying decision $S$ also likely affect the inclination of shifting the timing of the train journey $Y$. Furthermore, it is worth noting that $S=S(D)$ 
	implies that buying a supersaver ticket is conditional on receiving a non-zero discount. For this reason, non-treated subjects paying regular fares (with $D=0$) are not observed in our data. Yet, the outcome in our sample is defined relative to the behavior without treatment, as $Y$ indicates whether a has passenger has changed the timing of the trip because of a discount. This implies that $Y(0)=0$ by definition, such that the causal effect of some positive discount $d$ vs.\ no discount is $Y(d)-Y(0)=Y(d)$ is directly observable among observations that actually received $d$. However, it also appears interesting to investigate whether the demand shift effect varies across different (non-zero) discount rates $d$ $\in$ $(0,Q]$ to see whether the size matters. This is complicated by the fact that supersaver customers with different discount rates that are observed in our data might in general  differ importantly in terms of background characteristics also affecting the outcome, exactly because they bought their trip and were selected into the survey under non-comparable discount regimes. Our causal approach aims at tackling exactly this issue to establish customer groups that are comparable across discount rates in order to identify the effect of the latter.
	
	Based on the potential notation, we can define different causal parameters of interest. For instance, the average treatment effect (ATE) of providing discount levels $d$ vs.\ $d'$  (for $d\neq d'$) on outcome $Y$, denoted by $\Delta(d,d')$,
	corresponds to
	\begin{eqnarray}
		\Delta(d,d')=E[Y(d)-Y(d')].
	\end{eqnarray}
	Furthermore, the average partial effect (APE) of marginally increasing the discount level at $D=d$, denoted by ${\nabla}(d)$,  is defined as
	\begin{eqnarray}
		 {\nabla}(d)=\frac{\partial E[Y(d)]}{\partial d}.
	\end{eqnarray}
	Accordingly, ${\theta}=E[{\nabla}(D)]$ corresponds to the APE when marginally increasing the actually received discount of any individual (rather than imposing some hypothetical value $d$ for everyone).
	
	The identification of these causal parameters based on observable information requires rather strong assumptions. First, it implies that confounders jointly affecting $D$ and $Y$ can be controlled for by conditioning on observed characteristics. In our context, this appears plausible, as treatment assignment is based on variables related to demand (like weekdays or month), contingency schemes, capacity utilization, and supersaver tickets already sold - all of which is available in our data, as described in Section \eqref{data}. Second, identification requires that selection $S$ is as good as random (i.e., not associated with outcome $Y$) given the observed characteristics and the treatment, an assumption known as missing at random (MAR), see for instance \cite{Ru76b} and \cite{LittleRubin87}. However, the latter condition appears unrealistic in our framework, as our data lack important socio-economic characteristics likely affecting preferences and reservation prices for public transport, namely education, wealth, or income. For this reason, we argue that the ATE and APE among the individuals selected for the survey ($S=1$), i.e. \ conditional on buying a supersaver ticket, which are defined as
	\begin{eqnarray}
		\Delta_{S=1}(d,d')=E[Y(d)-Y(d')|S=1], {\theta}_{S=1}=E\left[{\nabla}(D)\vert S=1\right],
	\end{eqnarray}
	cannot be plausibly identified either. The reason is that if an increase in the discount rate induces some customers to buy a super saver ticket, then buyers with lower and higher discounts will generally differ in terms of their average reservation prices and related characteristics (as education or income), which likely also affect the demand-shift outcome $Y$.

	To tackle this sample selection issue, we exploit the fact that our data provide information on whether the supersaver customers would have purchased a ticket for this specific train trip also in the absence of any discount. Provided that the interviewees give accurate responses, we thus have information on $S(0)$, the hypothetical buying decision without treatment. Under the assumption that each customer's buying decision is weakly monotonic in the treatment in the sense that anyone purchasing a trip in a specific travel class (e.g., second class) without discount would also buy it for that class in the case of any positive discount, this permits identifying the group of always buyers. Importantly, we therefore define always buyers as those that would buy the trip not in a lower travel class (namely second rather than first class) without discount. For always buyers, $S(0)=S(d)=1$ for any $d>0$, such that their buying decision is always one and thus not affected by the treatment, implying the absence of the selection problem. In the denomination of \cite{FrangakisRubin02}, the always buyers constitute a so-called principal stratum, i.e., \ a subpopulation defined in terms of how the selection reacts to different treatment intensities. Therefore, sample selection bias does not occur within such a stratum, in which selection behavior is by definition homogeneous. For this reason, we aim at identifying the ATE and APE on the always buyers:
	\begin{eqnarray}
		\Delta_{S(0)=1}(d,d')&=&E[Y(d)-Y(d')|S(0)=1]=E[Y(d)-Y(d')|S(0)=S(d'')=1]\textrm{ for any }d'' \in (0,Q],\notag\\
		{\theta}_{S(0)=1}&=&E\left[{\nabla}(D)\vert S(0)=1 \right]=E\left[{\nabla}(D)\vert S(0)=S(d'')=1 \right],
	\end{eqnarray}
	where the respective second equality follows from the monotonicity of $S$ in $D$ which is formalized further below.
	
	
	\begin{figure}[!htp]
		\centering \caption{\label{figuresetup}  Causal framework}
		\includegraphics[scale=.5]{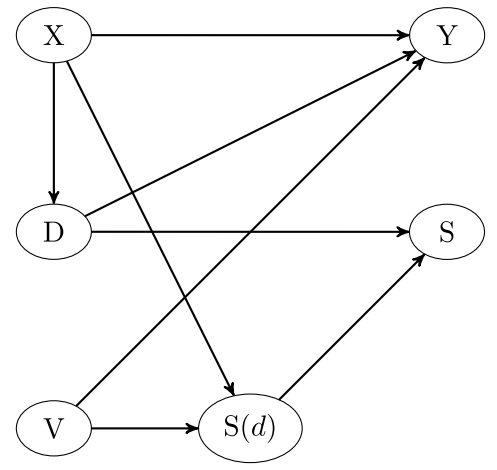}
	\end{figure}

	Figure \ref{figuresetup} provides a graphical illustration of our causal framework based on a directed acyclic graph, with arrows representing causal effects. Observed covariates $X$ that are related to demand are allowed to jointly affect the discount rate $D$ and the demand-shift outcome $Y$. $X$ may influence the potential purchasing decision under a hypothetical treatment $S(d)$, implying that buying a ticket given a specific discount depends on observed demand drivers like weekday, month, etc. Furthermore, unobserved socio-economic characteristics $V$ (like the reservation price) likely affect both $S(d)$ and $Y$. This introduces sample selection when conditioning on $S$, e.g.\ by only considering survey respondents ($S=1$). We also note that $S$ is deterministic in $D$ and $S(d)$ (as $S=S(D)$), even when controlling for $X$. This is the case because conditional on $S=1$, $D$ is associated with $V$, which also affects $Y$, thus entailing confounding of the treatment-outcome relation. A reason for this is for instance that buyers under higher and lower discounts are generally not comparable in terms of their reservation prices. In the terminology of \cite{Pearl00}, $S$ is a collider that opens up a backdoor path between $D$ and $Y$ through $V$. Theoretically, this could be tackled by jointly conditioning on the potential selection states under treatment values $d$ vs.\ $d'$ considered in the causal analysis, namely $S(d), S(d')$, as controls for the selection behavior. This is typically not feasible in empirical applications when only the potential selection corresponding to the actual treatment assignment is observed, $S=S(D)$. In our application, however, we do have information on $S(0)$ and can thus condition on being an always buyer under the mentioned monotonicity assumption.

	\subsection{Identifying Assumptions}
	
	We now formally introduce the identification assumptions underlying our causal analysis.\vspace{5pt}\newline
	\textbf{Assumption 1 (identifiability of selection under non-treatment):}\newline
	$S(0)$, is known for all subjects with   $S=1$. \vspace{5pt}\newline
	Assumption 1 is satisfied in our data in the absence of misreporting, as subjects have been asked whether they would have bought the train trip even in the absence of discount.
	\vspace{5pt}\newline
	\textbf{Assumption 2 (conditional independence of the treatment):}\newline
	${Y(d),S(d)}  \bot D | X$ for all $d \in (0,Q]$.\vspace{5pt}\newline
	By Assumption 2, there are no unobservables jointly affecting the treatment assignment on the one hand and the potential outcomes or selection states under any positive treatment value on the other hand conditional on covariates $X$. This assumption is satisfied if the treatment is quasi-random conditional on our demand-related factors $X$. Note that the assumption also implies that $Y(d)  \bot D | X, S(0)=1$ for all $d \in (0,Q]$.
	\vspace{5pt}\newline
	\textbf{Assumption 3 (weak monotonicity of selection in the treatment):}\newline
	$ \Pr(S(d)\geq S(d') | X)=1$ for all $d>d'$ and  $d,d'$ $\in$ $(0,Q]$. \vspace{5pt}\newline
	By Assumption 3, selection is weakly monotonic in the treatment, implying that a higher treatment state can never decrease selection for any individual. In our context, this means that a higher discount cannot induce a customer to not buy a ticket that would have been purchased under a lower discount. An analogous assumption has been made in the context of nonparametric instrumental variable models, see \cite{Imbens+94} and \cite{Angrist+96}, where, however, it is the treatment that is assumed to be monotonic in its instrument. Note that monotonicity implies the testable implication that $E[S-S(0)|X,S=1,D=d]=E[1-S(0)|X,S=1,D=d]$ weakly increases in treatment value $d$. In words, the share of customers that bought the ticket because of the discount must increase in the discount rate in our survey population when controlling for $X$.
	\vspace{5pt}\newline
	\textbf{Assumption 4 (common support):}\newline
	$f(d| X, S(0)=1)>0$ for all $d \in (0,Q]$.\vspace{5pt}\newline
	Assumption 4 is a common support restriction requiring that $f(d| X, S(0)=1)$, the conditional density of receiving a specific treatment intensity $d$ given $X$ and $S(0)=1$ (or conditional probability if the treatment takes discrete values), henceforth referred to as treatment propensity score, is larger than zero among always buyers for the treatment doses to be evaluated. This implies that the demand-related covariates $X$ do not deterministically affect the discount rate received such that there exists variation in the rates conditional on $X$.\\
	
	Our assumptions permit identifying the conditional ATE given $X$ (CATE), denoted by $\Delta_{X,S(0)=1}(d,d')=E[Y(d)-Y(d')|X,S(0)=1]$ for $d\neq d'$ and $d,d'$ $\in$ $(1,Q]$. To see this, note that
	\begin{eqnarray}
		\Delta_{X,S(0)=1}(d,d')&=&E[Y|D=d, X,S(0)=1]-E[Y|D=d',X,S(0)=1],\notag \\
		&=&E[Y|D=d, X,S(0)=1,S=1]-E[Y|D=d',X,S(0)=1,S=1],\label{cateid}
	\end{eqnarray}
	where the first equality follows from Assumption 2 and the second from Assumption 3, as monotonicity implies that asymptotically, $S=1$ if $S(0)=1$. Together with Assumption 1, which postulates the identifiability of $S(0)$, it follows that the causal effect on always buyers is nonparametrically identified, given that common support (Assumption 4) holds. If follows that the ATE among always buyers is identified by averaging over the distribution of $X$ given $S(0)=1,S=1$:
	\begin{eqnarray}
		\Delta_{S(0)=1}(d,d')=E[E[Y|D=d, X,S(0)=1,S=1]-E[Y|D=d',X,S(0)=1,S=1]|S(0)=1,S=1].
	\end{eqnarray}
	
	Furthermore, considering \eqref{cateid} and letting $d-d'\rightarrow 0$ identifies the conditional average partial effect (CAPE) of marginally increasing the treatment at $D=d$ given $X,S(0)=1$, denoted by ${\theta}_{X,S(0)=1}=E\left[{\nabla}(D) \vert X,S(0)=1 \right]$:
	\begin{eqnarray}
		{\theta}_{X,S(0)=1}=E \left[\frac{\partial E[Y\vert D=d, X,S(0)=1,S=1]}{\partial d} \big\vert X,S(0)=1,S=1\right].
	\end{eqnarray}
	Accordingly, the APE among always buyers that averages over the distributions of $X$ and $D$ is identified by
	\begin{eqnarray}
		{\theta}_{S(0)=1}=E\left[\frac{\partial E[Y\vert D=d, X,S(0)=1,S=1]}{\partial d} \big\vert S(0)=1,S=1\right].
	\end{eqnarray}
	
	Our identifying assumptions yield a testable implication if some personal characteristics (like customer's age) that affect $S(d)$ are observed, which we henceforth denote by $W$. In fact, $D$ must be statistically independent of $W$ conditional on $X,S(0)=1, S=1$ if $X$ is sufficient for avoiding any cofounding of the treatment-outcome relation. To see this, note that personal characteristics must by Assumption 2 not influence the treatment decision conditional on $X$. This statistical independence must also hold within subgroups (or principal strata) in which sample selection behavior (and thus sample selection/collider bias) is controlled for like the always buyers, i.e.\ conditional on $S(d), S=1$.
	
	\section{Estimation based on causal machine learning}\label{methods}

	In this section, we outline the causal machine learning approach used in our empirical analysis. Let $i$ $\in$ $\{1,....,n\}$ be an index for the different interviewees in our evaluation sample of size $n$ and $\{Y_i,D_i,X_i,W_i, S_i(0)\}$ denote the outcome, treatment, the covariates related to the treatment and the outcome, the observed personal characteristics, and the buying decision without discount of these interviewees that by the sampling design all satisfy $S_i=1$ (because they are part of the survey). Therefore, $Y_i$ represents customer $i$'s demand shift (rescheduling behavior) under customer $i$'s received discount rate $D_i$ relative to no discount. Our analysis assesses the causal effect of increasing discount rates on demand shifts among always buyers satisfying $S_i(0)=1$ while controlling for the selection into the survey and the non-random assignment of the treatment based on Assumptions 1 to 4 of Section \ref{ident}. We apply the causal forest (CF) approach of \cite{WagerAthey2018}, and \cite{AtheyTibshiraniWager2019} to estimate the CAPE and APE of the continuous treatment, as well as the double machine learning (DML) approach of \cite{Chetal2018} to estimate the ATE of a binary treatment of a discount $\geq 30\%$ vs.\ $<30\%$ in the sample of always buyers.\footnote{An introduction to causal machine learning techniques as well as coding examples are provided in the textbook by \cite{Huber2021causal}.}
	
	The CF adapts the random forest to the purpose of causal inference.  It is based on first running separate random forests for predicting the outcome $Y$ and the treatment $D$ as a function of the covariates $X$ using leave-one-out cross-fitting. The latter implies that the outcome or treatment of each observation is predicted based on all observations in the data but its own, in order to safeguard against overfitting bias. Second, the predictions are used for computing residuals of the outcomes and treatments, in which the influence of $X$ has been partialled out. Finally, a further random forest is applied to average over so-called causal trees, see \cite{AtheyImbens2016}, in order to estimate the CAPE.  The causal tree approach contains two key modifications when compared to standard decision trees. First, instead of an outcome variable, it is the coefficient of regressing the residual of $Y$ on the residual of $D$, i.e.\ the causal effect estimate of the treatment, that is to be predicted. Recursive splitting aims to find the largest effect heterogeneities across subsets defined in terms of $X$ to estimate the CAPE accurately. Secondly, within each subset, different parts of the data are used for estimating (a) the tree's splitting structure (i.e., \ the definition of covariate indicator functions) and (b) the causal effect of the treatment to prevent spuriously large effect heterogeneities due to overfitting.
	
	The CAPE estimate obtained by CF can be thought of as a weighted regression of the outcome residual on the treatment residual. The random forest-determined weight reflects the importance of a sample observation for assessing the causal effect at specific values of the covariates. After estimating the CAPE given $X$, the APE is obtained by appropriately averaging over the distribution of $X$ among the always buyers.  For implementing CAPE and APE estimation, we use the \textit{grf} package by \cite{Tibshiranietal2020} for the statistical software \textsf{R}. We set the number of trees to be used in a forest to 1000. We select any other tuning parameters like the number of randomly chosen covariates considered for splitting or the minimum number of observations per subset (or node) by the built-in cross-validation procedure.
	
	We also estimate the ATE among always buyers in our sample based on DML for a binary treatment defined as $\tilde{D}=I\{D\geq 0.3\}$, with $I\{\cdot\}$ denoting the indicator function that is equal to one if its argument is satisfied and zero otherwise. Furthermore, let $\mu_d(X)=E[Y|\tilde{D}=d, X,S(0)=1,S=1]$ denote the conditional mean outcome and $p_d(X)=\Pr(\tilde{D}=d| X, S(0)=1, S=1)$ the propensity score of receiving treatment category $d$ (with $d=1$ for a discount $\geq30\%$ and $d=0$ otherwise) in that population. Estimation is based on the sample analog of the doubly robust identification expression for the ATE, see \cite{Robins+94} and \cite{RoRo95}:
	\begin{eqnarray}\label{drselobs}
		\Delta_{S(0)=1}(1,0)&=&E\left[ \mu_1(X)-\mu_{0}(X)\right.\\
		&+&\left. \frac{(Y-\mu_1(X))\cdot \tilde{D}}{p_1(X)}-\frac{(Y-\mu_{0}(X))\cdot (1-\tilde{D})}{p_0(X)} \big| S(0)=1,S=1 \right].\notag
	\end{eqnarray}
	We estimate \eqref{drselobs} using the \textit{causalweight} package for the statistical software \textsf{R} by \citet{BodoryHuber2018}. As machine learners for the conditional mean outcomes $\mu_D(X)$ and the propensity scores $p_D(X)$ we use the random forest with the default options of the \textit{SuperLearner} package of \cite{vanderLaanetal2007}, which itself imports the \textit{ranger} package by \cite{WrightZiegler2017} for random forests. To impose common support in the data used for ATE estimation, we apply trimming threshold of $0.01$, implying that we drop observations with estimated propensity scores smaller than 0.01 (or 1\%) and larger than 0.99 (or 99\%) from our sample.

	\section{Descriptive analysis}\label{descriptives}

	Before discussing the results of our machine learning approaches, we first present some descriptive statistics for our data in Table \ref{t:table1}, namely the mean and the standard deviation of selected variables by always buyer status and binary discount category ($\geq 30\%$ and $<30\%$). We see that discounts and regular ticket fares of always buyers are on average lower than those of other customers. Another interesting observation is that in either discount category, we observe less leisure travelers among the always buyers than among other customers, which can be rationalized by business travelers responding less to price incentives by discounts. This is also in line with the finding that always buyers tend to purchase more second class tickets. More generally, we see non-negligible variation in demand-related covariates across the four subsamples defined in terms of buying behavior and discount rates. For instance, among always buyers, the total amount of supersaver tickets offered is on average larger in the higher discount category, while it is lower among the remaining clients. This suggests that neither the treatment nor being an always buyer is quasi-random, a problem we aim to tackle based on our identification strategy outlined in Section \ref{ident}. Concerning the demand-shift outcome, we see that always buyers change the departure time less frequently than others. With regard to upselling, we recognize that the relative amount of individuals upgrading their 2nd class to a 1st class ticket is the same for both discount categories, i.e. $\geq 30\%$ and $<30\%$.
	

	\begin{table}[h!]
		\begin{threeparttable}[b]
			\setlength{\tabcolsep}{0pt}
			\small
			\caption{Mean and standard deviation by discount and type} \label{t:table1}
			\begin{tabular*}{\textwidth}{ @{\extracolsep{\fill}} lcccc}
				\toprule
				discount & \multicolumn{2}{c}{$<$ 30\%} & \multicolumn{2}{c}{$\geq$ 30\%} \\
				always buyers & No & Yes & No & Yes  \\
				\midrule
discount & 0.21 & 0.19 & 0.57 & 0.53 \\
& (0.07) & (0.08) & (0.12) & (0.13) \\
regular ticket fare & 44.36 & 36.14 & 47.19 & 32.91 \\
& (29.38) & (25.47) & (30.14) & (23.78) \\
age & 47.22 & 47.68 & 45.59 & 48.77 \\
& (15.36) & (16.14) & (15.80) & (16.49) \\
gender & 0.51 & 0.55 & 0.53 & 0.59 \\
& (0.50) & (0.50) & (0.50) & (0.49) \\
diff. purchase travel & 3.42 & 3.23 & 7.72 & 7.19 \\
& ( 6.96) & ( 6.76) & (11.23) & (10.30) \\
distance & 136.49 & 127.86 & 126.15 & 116.76 \\
& (77.38) & (71.49) & (69.98) & (66.04) \\
capacity utilization & 35.51 & 39.19 & 26.46 & 33.15 \\
& (14.16) & (14.31) & (13.24) & (13.75) \\
seat capacity & 328.28 & 429.57 & 303.83 & 445.14 \\
& (196.19) & (196.10) & (185.42) & (188.54) \\
offer total & 33.95 & 44.10 & 70.97 & 98.34 \\
& (42.57) & (50.68) & (69.57) & (84.45) \\
sold total & 28.04 & 37.29 & 13.70 & 25.75 \\
& (41.92) & (50.31) & (36.37) & (53.67) \\
half fare travel ticket & 0.74 & 0.79 & 0.62 & 0.74 \\
& (0.44) & (0.40) & (0.49) & (0.44) \\
leisure & 0.77 & 0.69 & 0.82 & 0.76 \\
& (0.42) & (0.46) & (0.39) & (0.43) \\
class & 1.38 & 1.65 & 1.33 & 1.73 \\
& (0.48) & (0.48) & (0.47) & (0.44) \\
Swiss & 0.89 & 0.92 & 0.88 & 0.88 \\
& (0.31) & (0.28) & (0.33) & (0.32) \\
demand shift & 0.31 & 0.19 & 0.31 & 0.23 \\
& (0.46) & (0.40) & (0.46) & (0.42) \\
upselling & 0.49 & 0.00 & 0.49 & 0.00 \\
& (0.50) & (0.00) & (0.50) & (0.00) \\
obs. & 1151 & 2221 & 5529 & 3745 \\
				\bottomrule
			\end{tabular*}
			\begin{tablenotes}[flushleft]
				\footnotesize
				\item \textit{Notes: Regular ticket fare is in Swiss francs. `diff. purchase travel' denotes the difference between purchase and travel day. `Offer total' and `sold total' denote the total amount of supersaver tickets offered and the total amount of supersaver tickets sold respectively. }
			\end{tablenotes}
		\end{threeparttable}
	\end{table}

	As a further descriptive analysis, we apply the random forest, a predictive machine learner described in Appendix \ref{app_pred}, to investigate which variables among the covariates $X,W$ as well as the size of the discount $D$  importantly predict three outcomes of customer behavior. These outcomes indicate whether customers rescheduled their trip e.g.\ away from rush hours (demand shift), bought a first-class rather than a second-class ticket (upselling), or booked a trip otherwise not realized by train (additional trip), formally defined as $S_i-S_i(0)$. As $S_i=1$ is equal to one for everyone in our sample, the later outcome corresponds to  $1-S_i(0)$ and indicates whether someone has been induced purchase the ticket because of the discount, i.e.\ is not an always buyer.

For each of the outcomes, Table \ref{table:varimppred} presents the 30 most predictive covariates in the training set ordered in decreasing order according to a variable importance measure. The latter is defined as the total decrease in the Gini index (as a measure of node impurity in terms of outcome values) in a tree when including the respective covariate for splitting, averaged over all trees in the forest. The results suggest that trip- and demand-related characteristics like seat capacity, utilization, departure time, and distance are important predictors. Concerning personal characteristics, also customer's age appears to be relevant. Furthermore, also the treatment intensity $D$ has considerable predictive power. Interestingly, specific connections (defined by indicators for points of departure and destination) turn out to be less important characteristics conditional on the other covariates already mentioned. The resulting forecasts may for instance serve as a base for customer segmentation, e.g. into customer groups more and less inclined to book an additional trip or switch  classes or departure times.
	
	At the bottom of Table \ref{table:varimppred} we also report the correct classification rates for the three outcomes. While the accuracy in predicting a demand shift amounts to 58\%, which is somewhat better than random guessing but not particularly impressive, the performance is more satisfactory for predicting decisions about additional trips with an accuracy of 65\% and quite decent for upselling (82\%).\footnote{We note that when predicting upselling, we drop the variable `class', which indicates whether someone travels in the first or second class, and `seat capacity', which refers to the capacity in the chosen class, from the set of predictors. The reason is that upselling is defined as switching from second to first class, and therefore, the chosen class and the related seat capacities are actually part of the outcome to be predicted.} Tables \ref{table:varimppredlargetreat} and \ref{table:varimppredsmalltreat} in the Appendix present the predictive outcome analysis separately for subsamples with discounts $\geq 30\%$ and \ $<30\%$, respectively. In terms of which classes of variables are most predictive (trip- and demand-related characteristics, age, discount rate) and also in terms of accuracy, the findings are rather similar to those in Table \ref{table:varimppred}. In general, predictive machine learning appears useful for forecasting customer behavior in the context of demand for train trips, albeit not equally well for all outcomes of interest.

However (and in contrast to our causal machine learning approach for assessing the demand effects of discounts), the predictive analysis in this section does not yield the causal effects of the various predictors. In particular, our approach averages the predictions of the respective outcome over different levels of discounts $D$ and thus different customer types in terms of reservation price (related to $S(0)$) and unobserved background characteristics that likely vary with the treatment level. Therefore, we also perform the prediction analysis within subgroups defined upon the treatment level $D$ to see whether the set of important predictors is affected by the treatment intensity. To this end, we binarize the treatment such that it consists of two categories, namely (non-zero) discounts below 30\%, i.e.\ covering the treatment range $d$ $\in$ $(0, 0.3)$, and more substantial discounts of 30\% and more,  $d$ $\in$ $[0.3, 0.7]$, as $70\%$ is the highest discount observed in our data. The tables with the results of this approach are provided in Appendix \ref{furthertables}.

	\begin{landscape}
		\begin{table}[ht]
			\caption{ Predictive outcome analysis}\label{table:varimppred}
			{\scriptsize
				\begin{center}
					\begin{tabular}{cc|cc|cc}
						\hline \hline
						\multicolumn{2}{c|}{demand shift} &  \multicolumn{2}{c|}{upselling}  &  \multicolumn{2}{c}{additional trip}\\
						variable & importance   &  variable & importance  &  variable & importance \\
						\hline
						departure time &142.694&capacity utilization &295.924&seat capacity & 147.037 \\
						seat capacity &121.42&offer level B&188.861&D & 128.086 \\
						age &119.846&offer level C&149.911&age & 123.948 \\
						capacity utilization &119.606&D &132.095&departure time & 123.516 \\
						D &112.474&age&100.258&capacity utilization & 113.160 \\
						distance &112.143&departure time &98.909&distance& 101.730 \\
						offer level B &84.142&offer level A&93.303&offer level B & 84.989 \\
						diff. purchase travel &81.167&distance&87.319&diff. purchase travel & 80.236 \\
						offer level C &76.238&offer level D&85.408&offer level A & 78.507 \\
						offer level A &75.971&diff. purchase travel &62.841&offer level C & 77.097 \\
						number of sub-journeys &73.096&number of sub-journeys &55.978&number of sub-journeys & 69.443 \\
						offer level D &61.763&rel. sold level A&44.505&ticket purchase complexity & 64.498 \\
						ticket purchase complexity &57.071&ticket purchase complexity &41.819&offer level D & 56.888 \\
						rel. sold level A &51.377&offer level E&37.159&class & 51.456 \\
						rel. amount imputed values &42.222&rel. sold level B&34.462&rel. sold level A & 46.969 \\
						rel. sold level B &38.144&rel. amount imputed values &30.747&rel. amount imputed values & 38.869 \\
						adult companions &34.176&rel. sold level C&28.635&rel. sold level B & 36.484 \\
						rel. sold level C &28.201&adult companions &25.115&half fare & 35.785 \\
						offer level E &25.714&2019&18.88&adult companions & 34.446 \\
						gender &23.707&gender&18.47&halfe fare travel ticket & 28.465 \\
						amount purchased tickets &19.575&rush hour&17.448&gender & 25.419 \\
						German &18.659&Saturday&16.173&rel. sold level C & 24.679 \\
						travel alone &18.605&German&15.457&offer level E & 22.556 \\
						2019&18.082&leisure&15.304&leisure & 20.438 \\
						French &17.906&amount purchased tickets &15.112&no subscriptions & 19.793 \\
						saturday &17.487&travel alone &14.792&amount purchased tickets & 19.283 \\
						Friday &17.272&half fare &14.306&German & 19.119 \\
						peak hour &17.064&French&14.161&travel alone & 18.139 \\
						class &16.973&Thursday&13.413&2019& 17.192 \\
						leisure &16.892&scheme 20&13.411&French & 17.026 \\
						\hline
						correct prediction rate &  0.581  & &0.817 & &0.653 \\
						balanced sample size & 6962 & &6738 & &7000\\
						\hline
					\end{tabular}
					
				\end{center}
				
				\textit{Notes: `Offer level A', `offer level B', `offer level C', `offer level D' and `offer level E' denote the amount of supersaver tickets with discount A, B, C, D and E respectively. `Diff. purchase travel' denotes the difference between purchase and travel day. `Rel. sold level A', `rel. sold level B', `rel. sold level C' and `rel. sold level D' denote the relative amount of supersaver tickets offered with discount A, B, C and D respectively. The relative amounts are in relation to the seats offered. `No subscriptions' indicates not possessing any subscription. For predicting upselling, the covariates `class' and `seat capacity' are dropped.}
				\par
			}
		\end{table}
	\end{landscape}
	
	\section{Results}\label{resuls}
	
	\subsection{Testing the identification strategy}

	Before presenting the results for the causal analysis, we consider two different methods to partially test the assumptions underlying our identification strategy. First, we test Assumption 3 (weak monotonicity) by running the CF and DML procedures as well as a conventional linear regression in which we use buying an additional trip $(1-S(0))$, i.e.\ not being an always buyer, as outcome variable and $X$ as control variables in our sample of supersaver customers. The CF permits estimating the conditional change in the share of surveyed customers induced to buy an additional trip by modifying the discount rate $D$ given $X$ and $S=1$, i.e.\ $E\left[\frac{\partial E[(1-S(0))\vert D=d,X,S=1]}{\partial d}\big\vert X, S=1  \right]$, as well as the average thereof across $X$ conditional on sample selection,  $E\left[\frac{\partial E[(1-S(0))\vert D=d,X,S=1]}{\partial d}\big\vert S=1  \right]$. DML, on the other hand, yields an estimate of the average difference in the share of additional trips across the high and low treatment categories conditional on sample selection,  $E[E[(1-S(0))|D<0.3,X,S=1]-E[(1-S(0))|D\geq 0.3,X,S=1]|S=1]$. Finally, the linear regression of $(1-S(0))$ on $D$ and all $X$ in our sample tests monotonicity when assuming a linear model.
	
\begin{figure}[!htp]
		\centering \caption{\label{histmon} Monotonicity given $X$}
		\includegraphics[scale=.6]{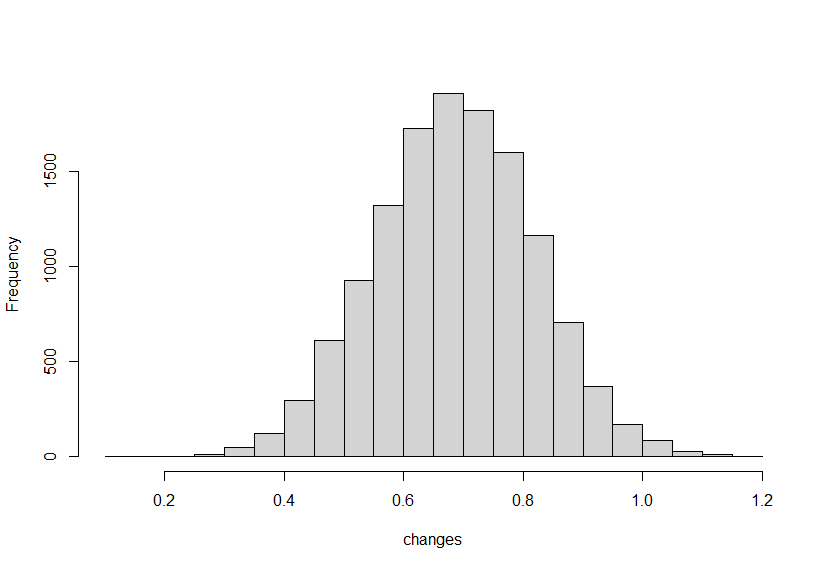}
	\end{figure}

	Table \ref{table:mon} reports the results, which do not provide any evidence against the monotonicity assumption. When considering the continuous treatment $D$, the CF-based estimate of $E\left[\frac{\partial E[(1-S(0))\vert D=d,X,S=1]}{\partial d}\big\vert S=1  \right]$ is highly statistically significant and suggests that augmenting the discount by one percentage point increases the share of customers otherwise not buying the ticket by 0.56 percentage points on average. Furthermore, any estimates of the conditional change $E\left[\frac{\partial E[(1-S(0))\vert D=d,X,S=1]}{\partial d}\big\vert X, S=1  \right]$ are positive, as displayed in the histogram of Figure \ref{histmon}, and 82.2\% of them are statistically significant at the 10\% level, 69.1\% at the 5\% level. Furthermore, the linear regression coefficient of  0.544 is highly significant. Likewise, the statistically significant DML estimate points to an increase in the share of additional trips by 18.4 percentage points when switching the binary treatment indicator from $D<0.3$ to $D\geq 0.3$.
	
	\begin{table}[h!]
		\caption{Monotonicity tests}\label{table:mon}
		{\footnotesize
			\begin{center}
				\begin{tabular}{r | cc c}
					\hline\hline
					& CF: average change & LR: coefficient & DML: $D\geq 0.3$ vs $D<0.3$ \\
					\hline
					change in ($1-S(0)$) & 0.564 & 0.544 & 0.184 \\
					standard error & 0.060 & 0.031 & 0.009 \\
					p-value & 0.000 & 0.000 & 0.000 \\
					trimmed observations &  &  & 1760 \\
					\hline
					number of observations & \multicolumn{3}{c}{12924}\\
					\hline
				\end{tabular}
			\end{center}
			\begin{tablenotes}[flushleft]
				\footnotesize
				\item \textit{Notes: `CF', `LR', and `DML' stands for estimates based on  causal forests, linear regression, and double machine learning, respectively. `trimmed observations' is the number of trimmed observations in DML when setting the propensity score-based trimming threshold to $0.01$. Control variables consist of $X$.}
			\end{tablenotes}
			\par
		}
	\end{table}

	We also test the statistical independence of $D$ and $W$ conditional on $X$ in our sample of always buyers, as implied by our identifying assumptions, see the discussion at the end of Section \ref{ident}. To this end, we run an linear regression of $D$ on $X$ and include one variable in $W$ at a time. Using a Wald test, we check whether the added variable improves the model. The results suggest that $W$ and $D$ are largely independent: the average p-value of the Wald test is 0.44, with only one regressor being marginally significant (a dummy for French-speaking with a p-value of 0.049). This result does not provide compelling statistical evidence that $W$ is associated with $D$ conditional on $X$.

	\subsection{Assessing the causal effect of discounts}
	
	Table \ref{table:eff} presents the main results of our causal analysis, namely the estimates of the discount rate's effect on the demand shift outcome, which is equal to one if the discount induced rescheduling the departure time and zero otherwise. We note that all covariates, i.e.\ both the trip- or demand-related factors $X$ and the personal characteristics $W$, are used as control variables, even though we have previously claimed that $X$ is sufficient for identification. There are, however, good reasons for including $W$ as well in the estimations. First, conditioning on the personal characteristics available in the data may reduce estimation bias if $X$ is - contrarily to our assumptions and to what our tests suggest - not fully sufficient to account for confounding. Second, it can also reduce the variance of the estimator, e.g.\ if some factors like age are strong predictors of the outcome. Third, having $W$ in the CF allows for a more fine-grained analysis of effect heterogeneity based on computing more `individualized' partial effects that (also) vary across personal characteristics.
	
	\begin{table}[ht]
		\caption{Effects on demand shift}\label{table:eff}
		{\footnotesize
			\begin{center}
				\begin{tabular}{r | c c c c}
					\hline\hline						
					& CF& DML& LR& PS-matching  \\
					& APE& ATE $D\geq 0.3$ vs $D<0.3$& coefficient& ATE $D\geq 0.3$ vs $D<0.3$ \\
					\hline
					effect& 0.161& 0.038& 0.082& 0.077\\
					standard error & 0.063& 0.010& 0.042& 0.082\\
					p-value &0.011& 0.000& 0.053& 0.352\\
					trimmed observations& &  151& & \\
					\hline
					number of observations & \multicolumn{4}{c}{5903}\\
					\hline
				\end{tabular}
			\end{center}
			\begin{tablenotes}[flushleft]
				\footnotesize
				\item \textit{Notes: `LR', `PS-Matching', `CF' and `DML' stands for estimates based on ordinary least squares, propensity score matching, causal forests, and double machine learning, respectively. `trimmed observations' is the number of trimmed observations in DML when setting the propensity score-based trimming threshold to $0.01$. Control variables consist of both $X$ and $W$.}
			\end{tablenotes}
			\par
		}
	\end{table}

Considering the estimates of the CF, we obtain an average partial effect (APE) of 0.161, suggesting that increasing the current discount rate among always buyers by one percentage point increases the share of rescheduled trips by 0.16 percentage points. This effect is statistically significant at the 5\% level. As a word of caution, however, we point out that the standard error is non-negligible such that the magnitude of the impact is not very precisely estimated. When applying DML, we obtain an average treatment effect (ATE) of 0.038 that is significant at the 1\% level, suggesting that discounts of 30\% and more on average increase the number of demand shifts by 3.8 percentage points compared to lower discounts, which is qualitatively in line with the CF. Furthermore, we find a decent overlap or common support in most of our sample in terms of the estimated propensity scores across lower and higher discount categories considered in DML, see the propensity score histograms in Appendix \ref{PSplots}. This is important as ATE evaluation hinges on the availability of observations with comparable propensity scores across treatment groups. Only 151 out of our 5903 observations are dropped due to too extreme propensity scores below 0.01 or above 0.99 (pointing to a violation of common support).\footnote{Our findings of a positive ATE remain robust when setting the propensity score-based trimming threshold to 0.02 (ATE: 0.042) or 0.05 (ATE: 0.045).} In summary, our results clearly point to a positive average effect of the discount rate on trip rescheduling among always buyers, which is, however, not overwhelmingly large.

For benchmarking purposes, we also consider the results of more `traditional' estimators, namely linear regression (using the continuous discount as for the CF) and nearest neighbor propensity score matching (using the binary discount definition as for the DML). To implement the latter, we estimate the propensity scores of having a larger discount by a probit specification, run nearest neighbor matching with the default values of the \textit{Matching} package by \cite*{Sekhon2011match}, and bootstrap the previous steps 2000 times to obtain a standard error for propensity score matching. Both linear regression and matching indicate that a higher discount leads to more consumers shifting their demand, yet, neither estimate is significant at the 5\% level, while the linear regression effect is still significant at the 10\% level. For propensity score matching, the lower significance goes together with a lower precision (or higher variance) of the effect estimate relative to DML. This might point to an overfitting issue related to a limited number of observations but a relatively large set of covariates to be included into the propensity score. Therefore, the data-driven selection of important covariates (rather than using any of them) by means of causal machine learning algorithms like DML may appear preferable to reduce the risk of overfitting and increase precision.

	\subsection{Effect heterogeneity}
	
\begin{figure}[!htp] \label{fig:eff}
		\centering \caption{\label{histcape} CAPEs}
		\includegraphics[scale=0.6]{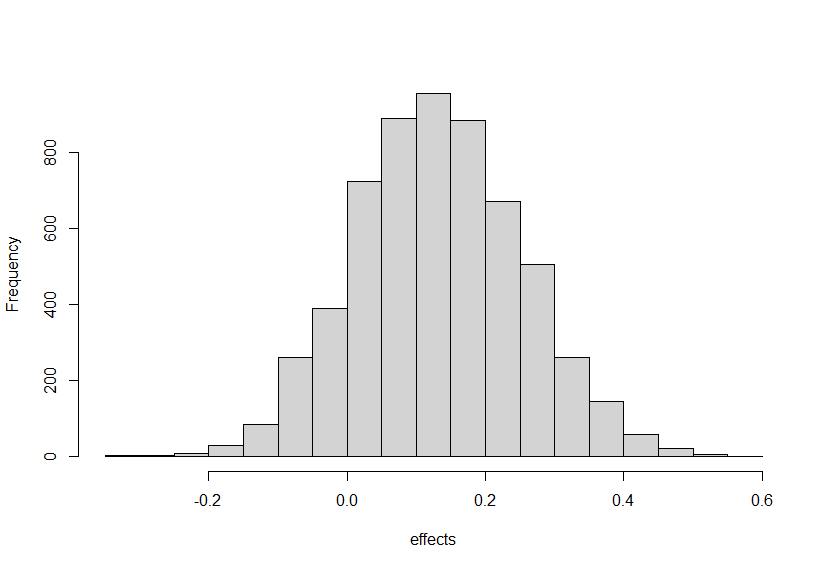}
	\end{figure}

	In this section, we assess the heterogeneity of the effects of $D$ on $Y$ across interviewees and observed characteristics. Figure  \ref{histcape} shows the distribution the CF-based conditional average effects (CAPE) of marginally increasing the discount rate given the covariates values of the always buyers in our sample (which are also the base for the estimation of the APE). While the CAPEs are predominantly positive, they are quite imprecisely estimated. Only 2.9\% and 0.8\% of the positive ones are statistically significant at the 10\% and 5\% levels, respectively. Further, only 0.1\% of the negative ones are statistically significant at the 10\% level. Yet, the distribution points to a positive marginal effect for most always buyers and also suggests that the magnitude of the effects varies non-negligibly across individuals.
	
	Next, we assess the effect heterogeneity across observed characteristics based on the CF results. First, we run a conventional random forest with the estimated CAPEs as the outcome and the covariates as predictors to assess the covariates' relative importance for predicting the CAPE, using the decrease in the Gini index as importance measure as also considered in Section \ref{descriptives}. Table \ref{table:varimp} reports the 20 most predictive covariates ordered in decreasing order according to the importance measure. Demand-related characteristics  (like seat capacity, utilization, departure time, and distance) turn out to be the most important predictors for the size of the effects, also customer's age has some predictive power. Similarly as for outcome prediction in  Section \ref{descriptives}, specific connections (characterized by points of departure or destination) are less important predictors of the CAPEs given the other information available in the data.
	
	\begin{table}[ht]
		\caption{ Most important covariates for predicting CAPEs}\label{table:varimp}
		{\scriptsize
			\begin{center}
				\begin{tabular}{cc}
					\hline \hline
					covariate & importance \\
					\hline
					seat capacity & 11.844 \\
					offer level C & 11.164 \\
					capacity utilization & 5.144 \\
					departure time & 5.122 \\
					distance & 4.287 \\
					offer level D & 4.015 \\
					class & 3.434 \\
					saturday & 2.933 \\
					age & 2.429 \\
					number of sections & 2.373 \\
					diff. purchase travel & 2.110 \\
					offer level A & 1.634 \\
					offer level B & 1.610 \\
					half fare & 1.524 \\
					scheme 17 & 1.496 \\
					half fare travel ticket & 1.373 \\
					rel. sold level B & 0.901 \\
					ticket purchase complexity & 0.847 \\
					leisure & 0.773 \\
					rel. sold level A & 0.770 \\
					\hline
				\end{tabular}
			\end{center}
			\begin{tablenotes}[flushleft]
				\footnotesize
				\item \textit{Notes: `Offer level A', `offer level B', `offer level C' and `offer level D' denote the amount of supersaver tickets with discount A, B, C and D respectively. `Rel. offer level A', `rel. offer level B' and `rel. offer level C' denote the relative amount of supersaver tickets offered with discount A, B and C. The relative amounts are in relation to the seats offered.}
			\end{tablenotes}
			\par
		}
	\end{table}

	While Table \ref{table:varimp} provides information on the best predictors of effect heterogeneity, it does not give insights on whether effects differ importantly and statistically significantly across specific observed characteristics of interest. For instance, one question relevant for designing discount schemes is whether (marginally) increasing the discounts is more effective among always buyers so far exposed to rather small or rather large discounts. Therefore, we investigate whether the CAPEs are different across our binary treatment categories defined by  $\tilde{D}$ (30\% or more and less than 30\%). To this end, we apply the approach of \cite{SemenovaChernozhukov2020} based on (i) plugging the CF-based predictions into a modified version of the doubly robust functions provided within the expectation operator of \eqref{drselobs} that is suitable for a continuous $D$ and (ii) linearly regressing the doubly robust functions on the treatment indicator $\tilde{D}$.
	The results are reported in the upper panel of Table \ref{table:effhetcat}. While the point estimate of $-0.104$ suggests that the demand shifting effect of increasing the discount is on average smaller when discounts are already quite substantial (above 30\%), the difference is far from being statistically significant at any conventional level.
	
	\begin{table}[ht]
		\caption{ Effect heterogeneity analysis}\label{table:effhetcat}
		{\footnotesize
			\begin{center}
				\begin{tabular}{cccc}
					\hline \hline
					& effect  & standard error & p-value \\
					\hline
					\emph{Discounts categories ($D\geq 0.3$ vs  $D<0.3$)}  & & & \\
					
					APE for $D<0.3$ (constant) & 0.209 & 0.089 & 0.019 \\
					Difference APE $D\geq 0.3$ vs  $D<0.3$ (slope coefficient)& -0.104 & 0.122 & 0.395 \\
					\hline
					\emph{Customer and travel characteristics}  & & & \\
					constant & -0.154 & 0.295 & 0.602 \\
					age & -0.002 & 0.004 & 0.556 \\
					gender & -0.022 & 0.129 & 0.866 \\
					distance & -0.000 & 0.001 & 0.697 \\
					leisure trip & 0.297 & 0.165 & 0.072 \\
					commute & 0.241 & 0.241 & 0.316 \\
					half fare travel ticket & 0.228 & 0.142 & 0.109 \\
					peak hours & 0.222 & 0.133 & 0.094 \\
					
					\hline
				\end{tabular}
			\end{center}
			\par
			
			\begin{tablenotes}[flushleft]
				\footnotesize
				\item \textit{Notes: Business trip is the reference category for the indicators `leisure trip' and `commute'.}
			\end{tablenotes}
		}
	\end{table}

	Using again the method of \cite{SemenovaChernozhukov2020}, we also investigate the heterogeneity among a limited and pre-selected set of covariates that appears interesting for characterizing customers and their travel purpose, namely age, gender, and travel distance, as well as indicators for leisure trip and commute (with business trip being the reference category), traveling during peak hours, and possession of a half fare travel tickets. As displayed in the lower panel of Table \ref{table:effhetcat}, we find no important effect heterogeneities across the age or gender of always buyers or as a function of travel distance conditional on the other information included in the regression, as the coefficients on these variables are close to zero. In contrast, the effect of demand shift is (given the other characteristics) substantially larger among always buyers with a half fare travel tickets and among commuters, however, neither coefficient is statistically significant at the 10\% level (even though the half fare coefficient is close).
	
	For leisure trips, the coefficient is even larger (0.297), suggesting that all other included variables equal, a one percentage point increase in the discount rate increases the share of rescheduled trips by 0.29 percentage points more among leisure travelers than among always buyers traveling for business. The coefficient is statistically significant at the 10\% level, even though we point out that the p-value does not account for multiple hypothesis testing of several covariates. This finding can be rationalized by leisure travelers being likely more flexible in terms of timing than business travelers. Also the coefficient on peak hours is substantially positive (0.222) and statistically significant at the 10\% level (again, without controlling for multiple hypothesis testing). This could be due to peak hours being the most attractive travel time, implying that costumers are more willing to reschedule their trips when being offered a discount within peak hours. We conclude that even though several coefficients appear non-negligible, statistical significance in our heterogeneity analysis is overall limited, which could be due to the (for the purpose of investigating effect heterogeneity) limited sample of several thousand observations.
	

	\section{Conclusion}\label{conclusion}
	
	In this study, we applied causal machine learning to assess the demand effects of discounts on train tickets issued by the Swiss Federal Railways (SBB), the so-called `supersaver tickets'. To this end, we analysed a unique data set that combines a survey of supersaver customers with rail trip- and demand-related information provided by the SBB. Our main result suggested that increasing the discount rate by one percentage point entails an average increase of 0.16 percentage points in the share of rescheduled trips among so-called always buyers (who would have travelled even without discount). We also found limited evidence for effect heterogeneity, e.g., that leisure travellers might be more responsive to discounts than other customer groups.
	
	Our study appears to be the first (at least for Switzerland) to provide empirical evidence on how discounts on train tickets affect customers' willingness to reschedule trips - an important information for designing discount schemes aiming at balancing out train utilization across time and reducing overload during peak hours.
	Even though the overall impact on the demand shifts on always buyers might not be as large as one could hope for, the causal forest pointed to the existence of customer segments that are likely more responsive and could be scrutinized when collecting a larger amount of data than available for our analysis. Furthermore, our empirical approach may also be applied to other countries or transport industries facing capacity constraints.  For instance, we would expect that in a setting with higher competition from alternative public transport modes like long distance bus services (not present in Switzerland), the impact of train discounts may well be different. More generally, our study can be regarded as a use case for how predictive and in particular causal machine learning can be fruitfully applied for business analytics and as decision support for optimizing specific interventions like discount schemes based on impact evaluation.

	\bibliographystyle{econometrica}
	\bibliography{research}
	
	\bigskip
	
	\renewcommand\appendix{\par
		\setcounter{section}{0}%
		\setcounter{subsection}{0}%
		\setcounter{table}{0}%
		\setcounter{figure}{0}%
		\renewcommand\thesection{\Alph{section}}%
		\renewcommand\thetable{\Alph{section}.\arabic{table}}}
	\renewcommand\thefigure{\Alph{section}.\arabic{subsection}.\arabic{subsubsection}.\arabic{figure}}
	\clearpage
	
	\begin{appendix}
		
		\numberwithin{equation}{section}
		\counterwithin{figure}{section}
		\noindent \textbf{\LARGE Appendices}

		{\small
		
		\section{Predicting customer behavior based on the random forest}\label{app_pred}
		
		The random forest is a nonparametric machine learner suggested by \cite{Breiman2001} for predicting outcomes as a function of observed variables, so-called predictors. Random forests rely on repeatedly drawing subsamples from the original data and averaging over the predictions in each subsample obtained by a decision tree, see \cite{Breimanetal1984}. The idea of decision trees is to recursively split the predictor space, i.e.\ the set of possible values of predictors, into a number of non-overlapping subsets (or nodes). Recursive splitting is performed such that after each split, a statistical goodness-of-fit criterion like the sum of squared residuals, i.e. \ the difference between the outcome and the subset-specific average outcome, is minimized across the newly created subsets. Intuitively, this can be thought of as a regression of the outcome on a data-driven choice of indicator functions for specific (brackets of) predictor values. At each split of a specific tree, only a random subset of predictors is chosen as potential variables for splitting in order to reduce the correlation of tree structures across subsamples, which together with averaging predictions over all subsamples reduces the estimation variance of the random forest when compared to running a single tree in the original data. Even when using an excessive number of splits (or indicator functions for predictor values) such that some of them do not importantly predict the outcome, averaging over many samples will cancel out those non-predictive splits that are only due to sampling noise. Forest-based predictions can be represented by smooth weighting functions that bear some resemblance with kernel regression, with the important difference that random forests detect good predictors in a data-driven way.
		
To predict the outcomes of customer behavior as discussed in Section \ref{descriptives}, we use the \textit{randomforest} package by \cite{breiman2018randomforest} for the statistical software \textsf{R} to implement the random forest based on growing 1,000 decision trees. To this end, we create three distinct datasets in which the values of the respective binary outcome are balanced, i.e.\ 1 (for instance, upselling) for 50\% and 0 (no upselling) for 50\% of the observations. We balance our data because we aim to train a model that predicts both outcome values equally well. Taking the demand shift outcome as an example, our data with non-missing covariate or outcome information contain 3481 observations with $Y=1$ and 9576 observations with $Y=0$. We retain all observations with $Y=1$ and randomly draw 3481 observations with $Y=0$ to obtain such a balanced data set. In the next step, we randomly split these 6962 observations into a training set consisting of 75\% of the data and a test set (25\%). In the training set, we train the random forest using the treatment $D$ and all covariates $X,W$ as predictors. In the test set, we predict the outcomes based on the trained forest, classifying e.g.\ observations with a demand shift probability $\geq 0.5$ as 1. We then compare the predictions to the actually observed outcomes to assess model performance based on the correct classification rate (also known as accuracy), i.e.\ the share of observations in the test data for which the predicted outcome corresponds to the actual one.}
			
			\section{Propensity score plots}\label{PSplots}
			
			
			\begin{figure}[!htp] \label{figPS1}
				\centering \caption{\label{fig:PS_Treat} Propensity score estimates in the higher discount category ( $D\geq 0.3$)}
				\includegraphics[scale=0.4]{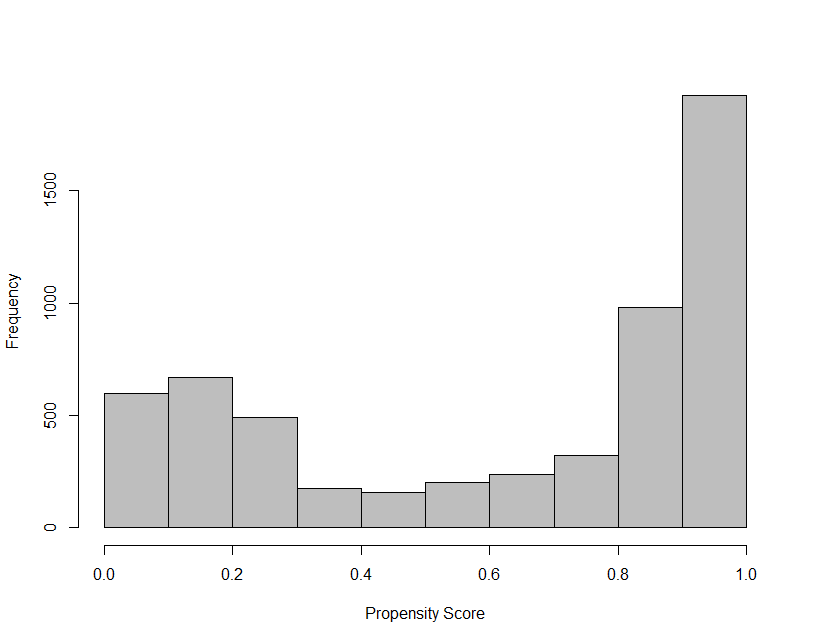}
			\end{figure}
			
			\begin{figure}[!htp] \label{figPS2}
				\centering \caption{\label{fig:PS_Control} Propensity score estimates in the lower discount category ($D<0.3$)}
				\includegraphics[scale=0.4]{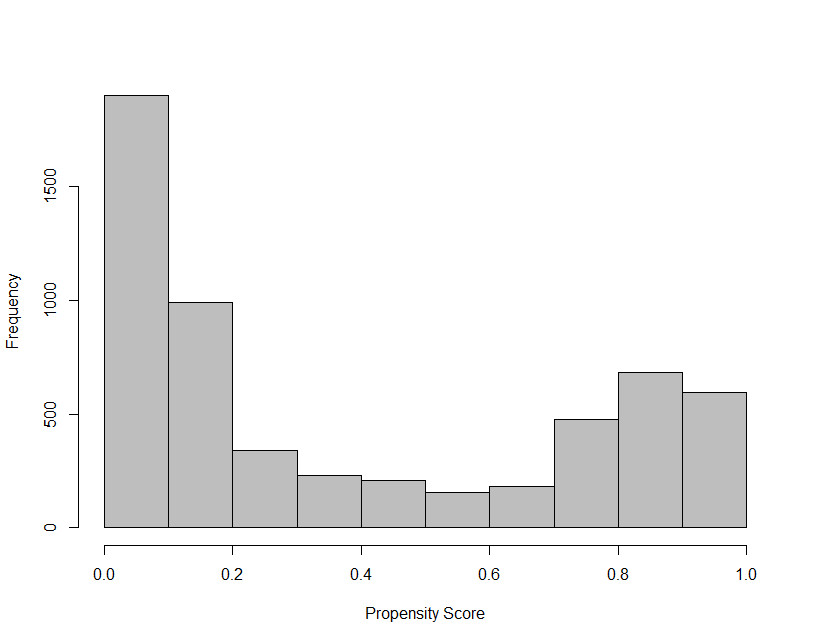}
			\end{figure}

			\section{Further tables }\label{furthertables}

			\begin{landscape}
				\begin{table}[ht]
					\caption{ Predictive outcome analysis, $D< 0.3$}\label{table:varimppredsmalltreat}
					{\scriptsize
						\begin{center}
							\begin{tabular}{cc|cc|cc}
								\hline \hline
								\multicolumn{2}{c|}{demand shift}  &  \multicolumn{2}{c|}{upselling} &  \multicolumn{2}{c}{additional trip} \\
								variable & importance    &  variable & importance &  variable & importance \\
								\hline
								departure time &37.33&capacity utilization &41.387&seat capacity & 25.342 \\
								seat capacity &27.871&offer level D&27.669&age & 21.639 \\
								capacity utilization &26.508&age&22.145&capacity utilization & 20.168 \\
								distance &26.31&D &19.077&distance& 18.970 \\
								age &26.223&offer level C &17.324&departure time & 18.527 \\
								D &25.08&departure time &16.538&D & 18.076 \\
								number of sub-journeys &17.403&distance&15.897&ticket purchase complexity & 16.637 \\
								offer level C &15.643&offer level B&15.696&offer level C & 12.085 \\
								diff. purchase travel &15.299&rel. sold level B &10.992&rel. sold level B & 11.709 \\
								ticket purchase complexity &15.116&number of sub-journeys &10.019&offer level D & 11.641 \\
								rel. sold level B &15.03&diff. purchase travel &9.573&number of sub-journeys & 11.347 \\
								offer level D &15.012&rel. sold level C&9.367&diff. purchase travel & 10.328 \\
								rel. sold level C &14.625&offer level A&7.857&offer level B & 10.185 \\
								offer level B &14.413&rel. sold level D&7.33&rel. sold level C & 8.993 \\
								rel. sold level A &11.856&ticket purchase complexity &7.319&rel. sold level A & 8.162 \\
								offer level A &11.329&offer level E&7.183&offer level A & 7.643 \\
								rel. amount imputed values &10.079&rel. sold level A&6.769&class & 7.381 \\
								rel. sold level D &9.625&rel. amount imputed values &5.422&rel. sold level D & 6.964 \\
								adult companions &8.503&adult companions&4.881&rel. amount imputed values & 6.189 \\
								offer level E &6.511&rush hour&4.143&adult companions & 5.785 \\
								gender &5.214&leisure&3.602&leisure & 5.398 \\
								leisure &5.154&gender&3.599&offer level E & 4.866 \\
								destination Geneva Airport &4.83&2019&3.597&gender & 4.692 \\
								departure Zuerich &4.736&travel alone&3.047&German & 3.801 \\
								class &4.598&Friday&2.92&halfe fare travel ticket & 3.686 \\
								travel alone &4.59&German&2.825&travel alone & 3.540 \\
								peak hour &4.545&French&2.479&French & 3.493 \\
								Friday &4.524&departure Zuerich&2.429&Friday & 3.419 \\
								German &4.522&destination Zuerich Airport&2.427&half fare & 3.163 \\
								amount purchased tickets &4.349&scheme 20&2.427&2019& 3.136 \\
								\hline
								correct prediction rate  &  0.555  &  & 0.772 & & 0.605\\
								balanced sample size  &    1642  &  &1140 & & 1202\\
								\hline
							\end{tabular}
						\end{center}
						\textit{Notes: `Diff. purchase travel' denotes the difference between purchase and travel day. `Rel. offer level A', `rel. offer level B', `rel. offer level C' and `rel. offer level D' denote the relative amount of supersaver tickets offered with discount A, B, C and D respectively. The relative amounts are in relation to the seats offered. `Offer level A', `Offer level B', `Offer level C', `Offer level D' and `Offer level E' denotes the amount of supersaver tickets with discount A, B, C, D and E respectively. `No subscription' indicates not possessing any subscription. For predicting upselling, the covariates `class' and `seat capacity' are dropped. }
						\par
				}\end{table}
			\end{landscape}
			
			\begin{landscape}

				\begin{table}[ht]
					\caption{ Predictive outcome analysis, $D\geq 0.3$}\label{table:varimppredlargetreat}
					{\scriptsize
						\begin{center}
							\begin{tabular}{cc|cc|cc}
								\hline \hline
								\multicolumn{2}{c|}{demand shift}  &  \multicolumn{2}{c|}{upselling} &  \multicolumn{2}{c}{additional trip}\\
								variable & importance   &  variable & importance   &  variable & importance\\
								\hline
								departure time &114&seat capacity &246.396&capacity utilization & 133.936 \\
								seat capacity &95.799&offer level B&178.212&age & 105.107 \\
								age &95.209&offer level C&127.327&departure time & 100.091 \\
								capacity utilization &89.422&D &100.618&capacity utilization & 97.889 \\
								distance &85.503&offer level A&88.947&distance & 85.647 \\
								D &80.447&Tageszeitinmin &82.658&D & 83.671 \\
								diff. purchase travel &69.276&age&78.886&offer level B & 73.399 \\
								offer level B &68.75&distance &73.885&offer level A & 69.936 \\
								offer level A &65.766&offer level D &72.452&diff. purchase travel & 67.823 \\
								offer level C &60.513&diff. purchase travel &55.321&offer level C & 64.689 \\
								number of sub-journeys &57.767&number of sub-journeys &48.622&number of sub-journeys & 58.100 \\
								offer level D &44.626&rel. sold level A &38.997&class & 54.857 \\
								ticket purchase complexity &44.434&ticket purchase complexity &31.991&ticket purchase complexity & 49.348 \\
								rel. sold level A &39.796&offer level E&27.041&offer level D & 46.685 \\
								rel. amount imputed values &32.144&rel. amount imputed values &25.586&rel. sold level A & 42.589 \\
								adult companions &25.925&adult companions &24.73&rel. amount imputed values & 35.411 \\
								rel. sold level B &20.536&rel. sold level B&17.099&half fare & 31.308 \\
								gender &18.629&gender&15.835&adult companions & 28.732 \\
								offer level E &17.521&2019&15.416&half fare travel ticket & 23.900 \\
								travel alone &15.15&Saturday&14.256&gender & 20.562 \\
								amount purchased tickets &14.939&amount purchased tickets &13.396&rel. sold level B & 20.036 \\
								German &14.855&rush hour&12.947&offer level E & 18.595 \\
								French &14.415&German&12.878&German & 16.257 \\
								2019&14.387&leisure&12.344&amount purchased tickets & 15.847 \\
								Sunday &13.821&travel alone &12.28&leisure & 15.500 \\
								destination Zuerich Airport &13.387&half fare&11.882&no subscription & 15.434 \\
								Saturday &13.378&Friday&11.646&travel alone & 15.132 \\
								class &13.27&scheme 20&11.559&Swiss & 14.951 \\
								half fare &13.258&French&11.477&Saturday & 14.613 \\
								rel. amount imputed values &13.048&Sunday&11.39&2019& 14.279 \\
								\hline
								correct prediction rate & 0.589  & &0.809 & & 0.629 \\
								balanced sample size &  5320 & &5598 & & 5798 \\
								\hline
							\end{tabular}
						\end{center}
						\textit{Notes: `Diff. purchase travel' denotes the difference between purchase and travel day. `Rel. offer level A', `rel. offer level B', `rel. offer level C' and `rel. offer level D' denote the relative amount of supersaver tickets offered with discount A, B, C and D respectively. The relative amounts are in relation to the seats offered. `Offer level A', `Offer level B', `Offer level C', `Offer level D' and `Offer level E' denotes the amount of supersaver tickets with discount A, B, C, D and E respectively. `No subscription' indicates not possessing any subscription. For predicting upselling, the covariates `class' and `seat capacity' are dropped. }
						\par
				}	\end{table}
			\end{landscape}

	\end{appendix}
\end{document}